\documentclass[twocolumn,epjc3]{svjour3}  
\smartqed  

\RequirePackage{graphicx}
\RequirePackage{dcolumn}
\RequirePackage{bm}
\RequirePackage{hyperref}
\usepackage[utf8]{inputenc}

\journalname{Eur. Phys. J. C}

\begin{document}

\title{Proton structure and hollowness from L\'evy imaging of $pp$ elastic scattering}

\author{T.~Cs\"org\H{o}\thanksref{e1,addr2,addr3} \and 
R.~Pasechnik \thanksref{e2,addr4,addr5,addr6} \and
A.~Ster\thanksref{e3,addr3} }

\thankstext{e1}{e-mail: tcsorgo@cern.ch}
\thankstext{e2}{e-mail: Roman.Pasechnik@thep.lu.se}
\thankstext{e3}{e-mail: Ster.Andras@wigner.mta.hu}

\institute{
EKE KRC, H-3200 Gy\"ongy\"os, M\'atrai \'ut 36, Hungary \label{addr2}
\and 
Wigner FK, H-1525 Budapest 114, P.O.Box 49, Hungary \label{addr3}
\and
Department of Astronomy and Theoretical Physics, 
Lund University, SE-223 62 Lund, Sweden \label{addr4}
\and
Nuclear Physics Institute ASCR, 25068 \v{R}e\v{z}, 
Czech Republic \label{addr5}
\and
Universidade Federal de Santa Catarina, C.P. 476, 
CEP 88.040-900, Florian\'opolis, 
SC, Brazil \label{addr6}
}


\maketitle

\begin{abstract}
\noindent
The recently developed L\'evy imaging met\-hod enables to extract an important physics information on hadron structure at high energies and ultra-low momentum transfers directly from elastic scattering data.
In this work, we employ such a model-independent method to probe the internal structure of the proton and quantify its inelasticity profile in the impact parameter space emerging in proton-proton collisions at the highest available energy of  $\sqrt{s} = 13$ TeV. The inelasticity profile function and its error band for the proton and its substructure have been reconstructed at different energies and the proton hollowness (or ``black-ring'') effect with beyond 5$\sigma$ significance has been found at 13 TeV.
\end{abstract}

The L\'evy stable distributions typically emerge in a description of dynamical systems' behaviour for which generalized central limit theorems apply, namely, as long as the resulting distribution for such a behavior is represented as a convolution of random elementary processes~\cite{Uchaikin:1999uz,Tsallis:1995zz}. Indeed, such distributions are very common in Nature and are often applied for probabilistic description 
of various correlated systems, anomalous diffusion, stochastic processes etc and not only in physics but also in other sciences~\cite{Uchaikin:1999uz}. In particle physics, a positively-definite modulus squared of a Fourier-trans\-formed source distribution often exhibits an approximate L\'evy stable shape emerging, in particular, in studies of two-particle Bose-Einstein correlation functions in high energy particle and heavy-ion physics \cite{Csorgo:1999wx,Csorgo:2003uv,Achard:2011zza,DeKock:2012gp,Novak:2016cyc,Adare:2017vig} and, more recently, in analysis of differential elastic scattering cross-sections at low momentum transfers~\cite{Csorgo:2018uyp}.

One of the most important and critical tests of Quantum Chromo Dynamics (QCD) in the infrared regime is provided by the ongoing studies of elastic differential hadron-hadron scattering cross section at various energies and momentum transfers. The characteristics of the elastic amplitude, its both real and imaginary parts, carry a plenty of information about the inner proton structure, the proton profile in the impact parameter space and its energy dependence, as well as about the properties of QCD exchange interaction at low momentum transfers. In this work, we apply the recently developed model-independent L\'evy imaging technique~\cite{Csorgo:2018uyp} for precision characterisation and extraction of relevant physics information about the proton structure and hence non-perturbative QCD directly from the available data on elastic cross sections at various energies. The results of our approach efficiently complement and provide an important guideline for the existing ongoing model-dependent efforts.

The first and most precise measurements of the total, elastic and differential cross sections of elastic $pp$ collisions has recently been performed by the TOTEM Collaboration at the Large Hadron Collider (LHC) 
at CERN at the highest energy frontier of $\sqrt{s} = 13$ TeV (for recent TOTEM publications, see Refs.~\cite{Antchev:2013gaa,Antchev:2017dia,Antchev:2017yns,Antchev:2018edk,Antchev:2018rec}). 
The large range in momentum transfer squared and very high precision of this set of data becomes a big challenge for a statistically acceptable description that is necessary for a reliable extraction of any physics information from such data with an appropriate statistical significance. A correct theoretical interpretation of the LHC data, together with the lower-energy Tevatron and ISR data, is a subject of intense debates and ongoing research development, see e.g. Refs.~\cite{Khoze:2017swe,Khoze:2018bus,Broilo:2018els,Martynov:2018nyb,Khoze:2018kna,Troshin:2018ihb} while a proper extraction of model-independent ``data-driven'' characteristics of the proton as it appears in a high-energy scattering is still lacking.

The L\'evy technique for proton imaging at the femtometer scale provides not only the impact parameter dependent inelasticity profile of the proton at various energies, but also the inelasticity profile of its internal substructure. In this work, such profiles and their energy dependence are reconstructed from the top LHC collision energy down to the lowest ISR energy, together with their error bars, for the first time.
At $\sqrt{s}= 13$ TeV, we find a statistically significant evidence for a hollowness (or ``black-ring'') effect that may fundamentally change the standard picture of $pp$ collisions at asymptotically 
large energies.

The TOTEM Collaboration has established~\cite{Antchev:2018edk,Antchev:2015zza} that at low values of the four-momentum transfer squared $t= (p_1 - p_3)^2 $ the differential elastic cross-section for $pp$ collisions differs from a conventionally assumed naive exponential form, $d\sigma/dt = A \exp(- B|t|)$, -- a subtle but significant deviation. A minimal way to parametrize such a deviation is to introduce
a single parameter $\alpha$ into the exponent as $d\sigma/dt = A \exp\left[- (R^2 |t|)^{\alpha}\right]$, to the leading order. This is the so-called stretched exponential distribution. It corresponds to the Fourier-transform of a symmetric L\'evy-stable source distribution~\cite{Uchaikin:1999uz}. This approximation with $\alpha = 0.9$ gives a statistically acceptable description of $pp$ elastic scattering at low-$t$ at $\sqrt{s} = $ 23.5 GeV to 62.5 GeV, while at the LHC energies higher order L\'evy expansion terms become relevant for description of the low-$t$ behaviour~\cite{Csorgo:2018uyp}. At larger $|t|$, deviation of the data from 
the L\'evy stable source distribution becomes even more pronounced in a vicinity of the dip-and-bump structure before turning back to a simple stretched exponential behaviour again (but with a different value 
of $R$) at large $|t|$ beyond the secondary maximum.

In momentum representation, the elastic differential cross-section
\begin{eqnarray}
\frac{d\sigma}{dt} = \frac{1}{4\pi}|T_{el}(\Delta)|^2 \,, \qquad \Delta=\sqrt{|t|}\, .
\label{e:dsigmadt-Tel}
\end{eqnarray}
is related to the modulus of the elastic amplitude $T_{el}(\Delta)$. To quantify the deviations of this observable from a symmetric L\'evy source distribution, 
we utilize an orthonormal series expansion for the complex-valued elastic scattering amplitude $T_{el}(\Delta)$ in terms of the L\'evy 
polynomials~\cite{Csorgo:2018uyp} as follows:
\begin{eqnarray}
T_{el}(\Delta) & = & 
    i\sqrt{4\pi A}\, e^{- z^{\alpha}/2}  
                \!  \left(\!
                c_0+\sum_{i = 1}^\infty c_i l_i (z|\alpha) \! \right)_{z = \Delta^2 R^2} \!\!\!\!\!\!\!\!\!\! \,
        \label{e:Tel}
\end{eqnarray}
where $R$ is the L\'evy scale parameter, $c_i = a_i + i b_i$ are the complex expansion coefficients, and $l_j(z|\alpha)$ corresponds to the normalized L\'evy polynomial 
of order $j$. For simplicity, we start this expansion with $c_0 = a_0 + i b_0 = 1$ as a possible value of $a_0$ gets absorbed to the constant normalization $A$ while 
a vanishing value of $b_0$ consistent with zero was found in data fits at all considered energies. 
\begin{figure}[ht]
        \includegraphics[scale=0.43]{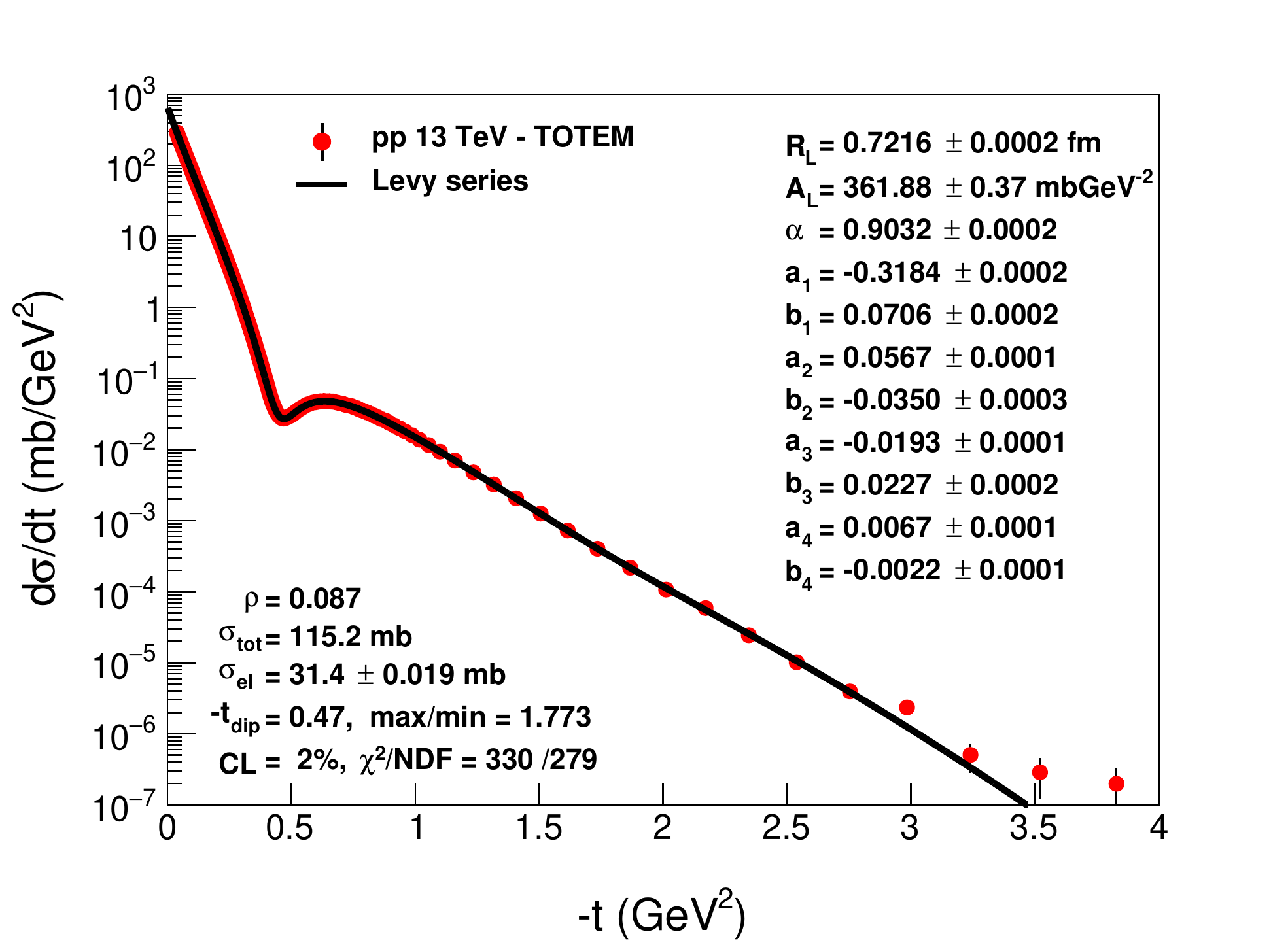}
        \caption{(Color online)
                Eq.~(\ref{e:dsigmadt-Tel}) fitted to the recent LHC TOTEM $d\sigma/dt$ data at $\sqrt{s} = 13$ TeV. 
}
        \label{fig:levyfit-13TeV}
\end{figure}

The normalized L\'evy polynomials $l_j(z|\alpha)$ in Eq.~(\ref{e:Tel}) are given 
\begin{eqnarray}
        l_j(z\, |\, \alpha) & = & D^{-\frac{1}{2}}_{j} D^{-\frac{1}{2}}_{j+1} 
        L_j(z\, |\, \alpha) \,, \quad \mbox{\rm for}\quad j\ge 0 \, .
    \label{e:lj}
\end{eqnarray}
in terms of the the unnormalized L\'evy polynomials $L_i(z\,|\,\alpha)$ (with $L_0(z\,|\,\alpha) = 1$)~\cite{Novak:2016cyc}
\begin{eqnarray}
L_1(z\,|\,\alpha)  &=& 
       \det\left(\begin{array}{c@{\hspace*{8pt}}c}
     \mu_{0}^{\alpha} & \mu_{1}^{\alpha}  \\ 
     1 & z \end{array} \right) \,, \\
 L_2(z\,|\,\alpha)  &=& 
       \det\left(\begin{array}{c@{\hspace*{8pt}}c@{\hspace*{8pt}}c}
     \mu_{0}^{\alpha} & \mu_{1}^{\alpha} & \mu_{2}^{\alpha} \\ 
     \mu_{1}^{\alpha} & \mu_{2}^{\alpha} & \mu_{3}^{\alpha}  \\ 
     1 & z & z^2 \end{array} \right), \, ... \,
          \mbox{\rm etc} . 
\end{eqnarray}
and the Gram-determinants $D_j \equiv D_j(\alpha)$ ($D_0(\alpha) \equiv  1$)
\begin{eqnarray}
D_1(\alpha)  &=&  \mu_{0}^{\alpha} \,, \quad
D_2(\alpha)   = 
     \det\left(\begin{array}{c@{\hspace*{8pt}}c}
     \mu_{0}^{\alpha} & \mu_{1}^{\alpha}  \\ 
     \mu_{1}^{\alpha} & \mu_{2}^{\alpha} 
     \end{array} \right)\,, \, ...  \, \mbox{\rm etc} \,. \label{e:gram} \\
    \mu_{n}^{\alpha} &=& \int_0^\infty dz\;z^{n} e^{-z^\alpha} = \frac{1}{\alpha}\,\Gamma\left( \frac{n+1}{\alpha}\right) \,.
\end{eqnarray}
Here, $\Gamma(x) = \int_0^\infty dz\;z^{x-1}e^{-z}$ stands for Euler's gamma function.
The total and elastic cross-sections are then represented as follows
\begin{eqnarray}
\null & \null &
\sigma_{\rm tot}  \equiv   2\,{\rm Im}\, T_{el}(0)   =
2\,\sqrt{4\pi A}\,
        \left(1 + \sum_{i = 1}^\infty a_i l_i (0|\alpha) \right) \!\! ,  \null \null
    \label{e:sigmatot} \\
\null & \null &
    \sigma_{\rm el}  \equiv  \int_0^\infty \!\!\! d|t| \frac{\mbox{\rm d}\sigma}{\mbox{\rm d} t}   =   
    \frac{A}{R^2} \! \left[\frac{1}{\alpha}\Gamma\left(\frac{1}{\alpha}\right)\! + \! \sum_{i = 1}^\infty (a_i^2 + b_i^2) \right] \!\! .   \null  \null
    \label{e:sigmael}
\end{eqnarray}

The momentum transfer squared distribution of elastic scattering provides images of the scattered particles in impact parameter 
or $\mathbf b$ space. For elastic $pp$ collisions,
\begin{eqnarray}
        t_{el}(b) &=& \frac{1}{2\pi} \int J_0(\Delta\,b)\,T_{el}(\Delta)\,\Delta\, d\Delta =i\left[ 1 - e^{-\Omega(b)} \right] \,, \nonumber \\
        P(b) &=& 1-\left|e^{-\Omega(b)}\right|^2 \, , \label{tel-b}
\end{eqnarray}
where $ \Delta\equiv|\mathbf{\Delta}|$,  $b\equiv|\mathbf{b}|\, $, $t_{el}(b)$ is the impact parameter dependent elastic 
amplitude represented in an eikonal form in terms of the complex opacity (eikonal) function $\Omega(b) $, $J_0(x)$ is the zeroth-order Bessel 
function of the first kind, and $P(b)$ is the inelasticity (or shadow) profile function describing the proton image. The fundamental results 
of multiple diffraction theory~\cite{Glauber:1984su}, together with the L\'evy expansion of the elastic scattering amplitude $T_{el}(\Delta)$ (\ref{e:Tel}), 
make it possible to perform a high precision optical imaging of elastic $pp$ scattering and to obtain images of the protons and its internal structures 
at the femtometer length scale.
\begin{figure}[tb]
 \includegraphics[scale=0.55]{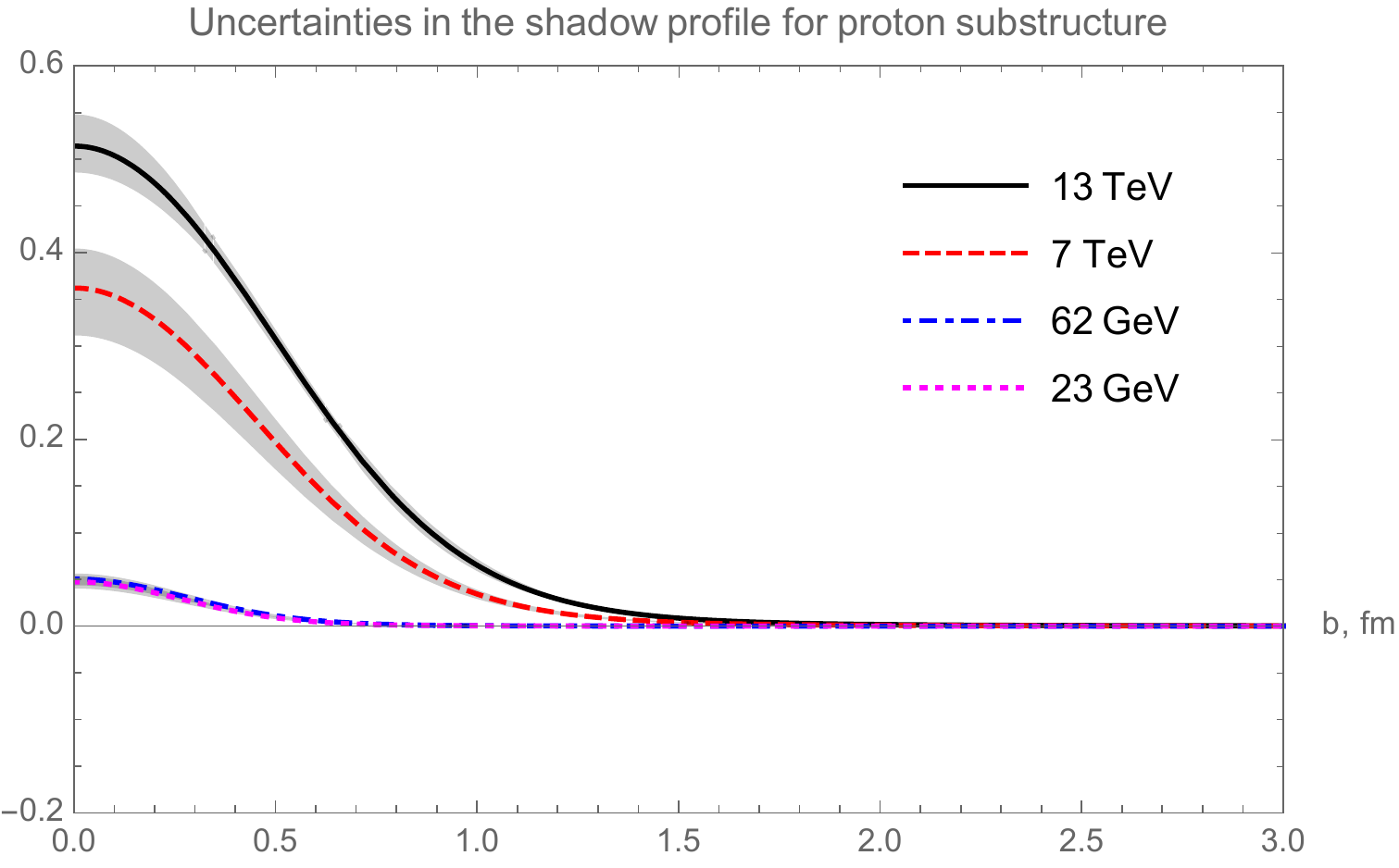}
        \caption{(Color online)
     The inelasticity profile $P(b)$ of proton substructure at various $\sqrt{s}$ (with error bars). Here, only the secondary 
     diffractive cone beyond the dip-and-bump structure is concerned.
        }
        \label{fig:shadow-subs}
\end{figure}
\begin{figure}[tb]
       \includegraphics[scale=0.55]{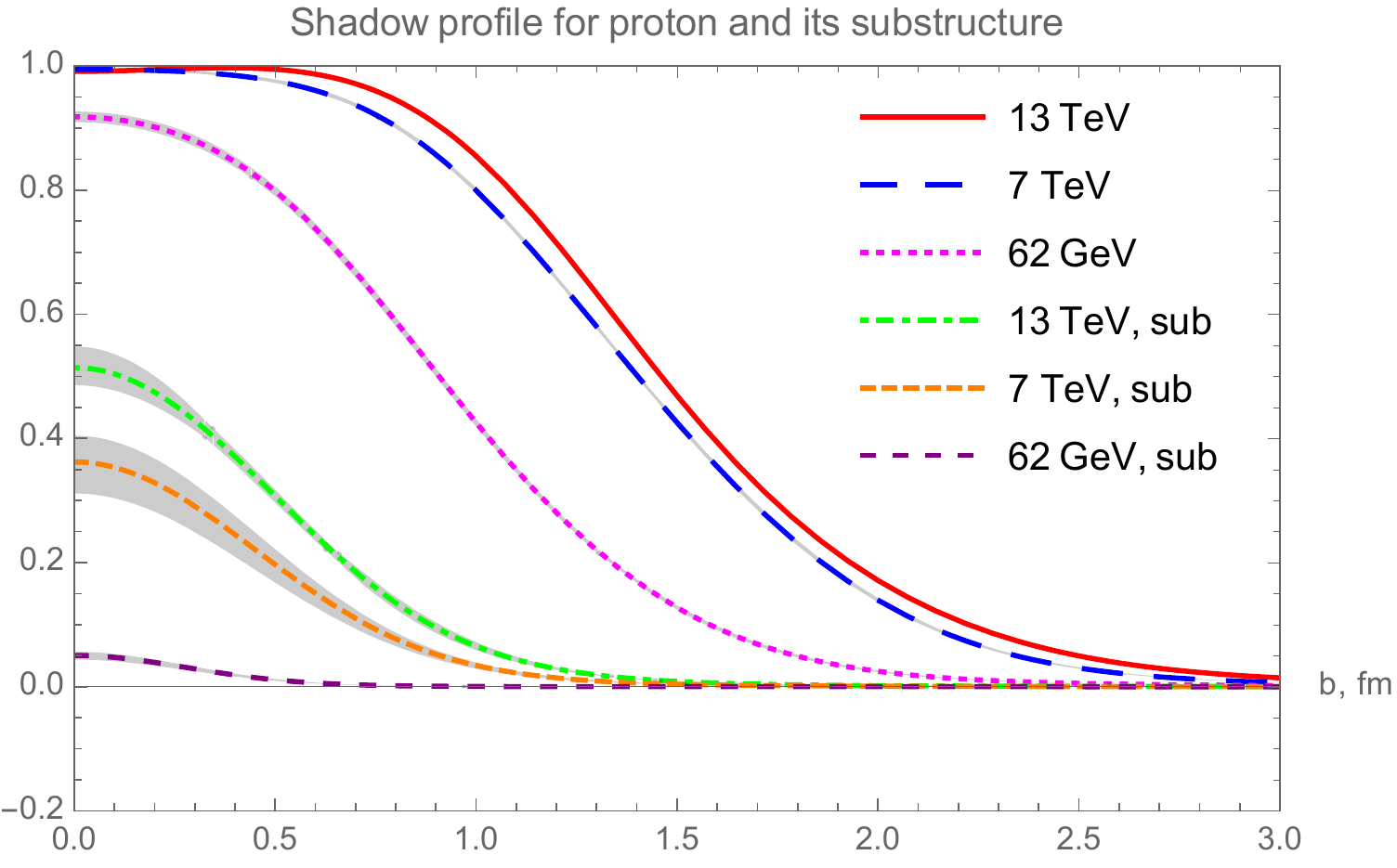} 
        \caption{(Color online)
        The inelasticity profile $P(b)$ of the proton and its substructure at various collision energies, including the error bars.
        }
        \label{fig:shadow-all}
\end{figure}
\begin{table}[t]
    \centering
    \begin{tabular}{c|c|c|c|c|c} %
       $\sqrt{s}$ (GeV)   & 23.5 & 30.7 & 44.7 & 52.8 & 62.5 
       \\ \hline
         A (mb GeV$^2$) & 0.003 $\pm$ 0.001 &
           0.006 $\pm$ 0.001 & 
               0.004 $\pm$ 0.001 & 
                    0.005 $\pm$ $\pm$ 0.001& 
                        0.004 $\pm$ 0.001 \\
         R (fm) & 0.30 $\pm$ 0.01 & 
            0.32 $\pm$ 0.01& 
                0.31 $\pm$ 0.01 & 
                    0.32 $\pm$ 0.01& 
                        0.31 $\pm$ 0.01  \\
         $\chi^2/{\rm NDF}$ & 22.7 / 21 & 
            16.3 / 21 & 
                19 / 21 & 
                    13.9 / 21 & 
                        12.2 / 21 \\
         CL [\%] &  36 & 
            75 & 
                59 & 
                   87 & 
                        94  \\ \hline
         $\sigma^{\mbox{\rm sub}}_{\mbox{\rm tot}}$ (mb)  & 0.24 $\pm$ 0.03 & 
                0.35 $\pm$ 0.03    & 
                       0.28 $\pm$ 0.02 & 
                            0.32 $\pm$ 0.01 & 
                                0.30 $\pm$ 0.03  \\ 
 $\sigma^{\mbox{\rm sub}}_{\mbox{\rm el}}$ ($\mu$b)  & 
        1.3 $\pm$ 0.3 & 
                2.4 $\pm$ 0.4    & 
                       1.7 $\pm$ 0.2 & 
                            2.1 $\pm$ 0.2 & 
                                1.9 $\pm$ 0.3  %
    \end{tabular}
    \caption{ Parameters of the L\'evy stable source distribution, $d\sigma/dt = A \exp\big[-(|t|R^2)^{\alpha}\big]$, 
    characterising the proton substructure at large $|t|$ ($1.5 \le -t \le 3$ GeV$^2$) corresponding to the secondary diffractive 
    cone beyond the dip-and-bump structure, and at various ISR collision energies, for $\alpha = 0.9$.
      }
    \label{t:sub-ISR}
\end{table}
\begin{table}[t]
    \centering
    \begin{tabular}{c|c|c} %
       $\sqrt{s}$ (TeV)   & 7 & 13   
       \\ \hline
         A (mb GeV$^2$) & 1.4 $\pm $ 0.4 &
           4.0 $\pm$ 0.3 \\
         R (fm) & 0.49 $\pm$ 0.01 & 
            0.51 $\pm$ 0.01 \\
         $\chi^2/{\rm NDF}$ & 4.9 / 10 & 
            16.9 / 8 \\
         CL [\%] &  90 & 
           3  \\ \hline
         $\sigma^{\mbox{\rm sub}}_{\mbox{\rm tot}}$ (mb)  & 5.2  $\pm$ 0.7  & 
                8.9 $\pm$ 0.30    \\ 
         $\sigma^{\mbox{\rm sub}}_{\mbox{\rm el}}$ (mb)  & 0.24  $\pm$ 0.06  & 
                0.63 $\pm$ 0.04   
    \end{tabular}
    \caption{The same as in Table~\ref{t:sub-ISR} but at LHC energies.
    }
    \label{t:sub-LHC}
\end{table}

The power of the L\'evy imaging technique is demonstrated in Fig.~\ref{fig:levyfit-13TeV} where it is employed to describe the precision TOTEM data on elastic $pp$ scattering at the highest accessible 
energy $\sqrt{s} = $ 13 TeV~\cite{Antchev:2018edk} with confidence level CL = 2 \%. Given such an unprecedented precision of these data spanning ten orders of magnitude, the L\'evy expansion thus 
represents them in a statistically acceptable manner which still remains a big challenge for alternative model dependent approaches.

As usual, we notice a nearly exponential diffractive cone at low $|t|$ that is followed by a dip, then by a subsequent maximum and a secondary diffractive cone at large $t$. The maximal $t$ value accessed by the measurement $-t_{\rm max} \approx 4 $ GeV$^2$ provides an estimate for a minimal spatial resolution of $\hbar/\sqrt{-t_{\rm max}} \approx 0.1 $ fm. Remarkably, such a high resolution achieved at LHC energies enables to resolve smaller structures inside the colliding protons. Likewise, the L\'evy imaging technique is also useful to characterise the results of the earlier measurements of elastic $pp$ and $p\bar p$ 
scattering including those at ISR and Tevatron energies and hence to probe the proton structure in the full range of $\sqrt{s}$ from 23.5 GeV up to 13 TeV. Full $|t|$ range fits are used to obtain the shadow profiles $P(b)$ while fits in the large $|t|$ regions are used to get the shadow profiles for the substructure inside the protons. These fits are detailed in Appendix A as well as Appendix C of Ref.~\cite{Csorgo:2018uyp}. In this Letter, we present the error-bands around the reconstructed shadow profile functions and, hence, evaluate the significance of these results for the first time, as indicated on Figs.~\ref{fig:shadow-subs} and~\ref{fig:shadow-all}, with the fit results summarized in Tables~\ref{t:sub-ISR} and~\ref{t:sub-LHC}.

The proton substructure found at the energies of ISR appears to be rather faint, with $P(b=0)$ being about 0.05 that weakly depends on energy. However, at TeV energy scale (7 and 13 TeV) one observes
a much larger and significantly darker substructure, again not strongly evolving from 7 TeV to 13 TeV as demonstrated in Fig.~\ref{fig:shadow-subs}. The parameters of the L\'evy stable source distribution, 
$d\sigma/dt = A \exp\big[-(|t|R^2)^{\alpha}\big]$, characterising the secondary diffractive cone at large $|t|$, and hence describing the corresponding substructure, are given with their error bars 
in Tables~\ref{t:sub-ISR} and \ref{t:sub-LHC} at ISR and LHC energies, respectively. Besides, we provide the contributions of the scattering off such a substructure to the total and elastic 
cross-sections, $\sigma_{\rm tot}^{\rm sub}$ and $\sigma_{\rm el}^{\rm sub}$, found from Eqs.~(\ref{e:sigmatot}) and (\ref{e:sigmael}) with $(a_i,b_i) = (0,0)$, respectively. A comparison of the inelasticity 
profiles of the proton and its substructure at different $\sqrt{s} = $ 13 TeV, 7 TeV and 62 GeV is illustrated in Fig.~\ref{fig:shadow-all} together with error-bands.

The inelasticity (or shadow) profile function $P(b)$ of the protons undergoes a qualitative change at around  $\sqrt{s} \approx 7 $ TeV collision energy. At small values of the impact parameter $b$ a $P(b) \approx 1$ plateaux develops, which becomes depressed at larger energies, and a shallow minimum is formed near $b = 0$. Such a dip or hollowness may  correspond to $\sigma_{\rm el} \ge \sigma_{\rm tot}/4/ \big(1 + \rho_0^2\big)$~\cite{Csorgo:2019fbf}. The existence of such a hollow in high energy $pp$ scattering  is a hotly debated, current topic in the literature. We recommend Refs.~\cite{Troshin:2007fq,Fagundes:2011hv,Dremin:2013qua,Alkin:2014rfa,Anisovich:2014wha} for early papers as well as Refs.~\cite{RuizArriola:2016ihz,Albacete:2016pmp,Broniowski:2017aaf,Troshin:2017ucy,Dremin:2018urc,Broniowski:2018xbg,Petrov:2018wlv} and Ref.~\cite{Csorgo:2019fbf} for more recent theoretical discussions as well as an experimental outlook on this fundamental nature of $pp$ scattering at LHC and asymptotic energies. The maximal value of $P(s,b=0) \approx 1/(1 + \rho_0^2) = 1 -H $ at $\sigma_{\rm el}(s) \approx \sigma_{\rm tot}(s)/4/(1+\rho^2)$ seems to be rather independent of the detailed $b$-dependent  shape of the inelastic collisions, see for example Refs.~\cite{Broniowski:2018xbg,Csorgo:2019fbf}. 
\begin{figure}[ht]
        \includegraphics[scale=0.3]{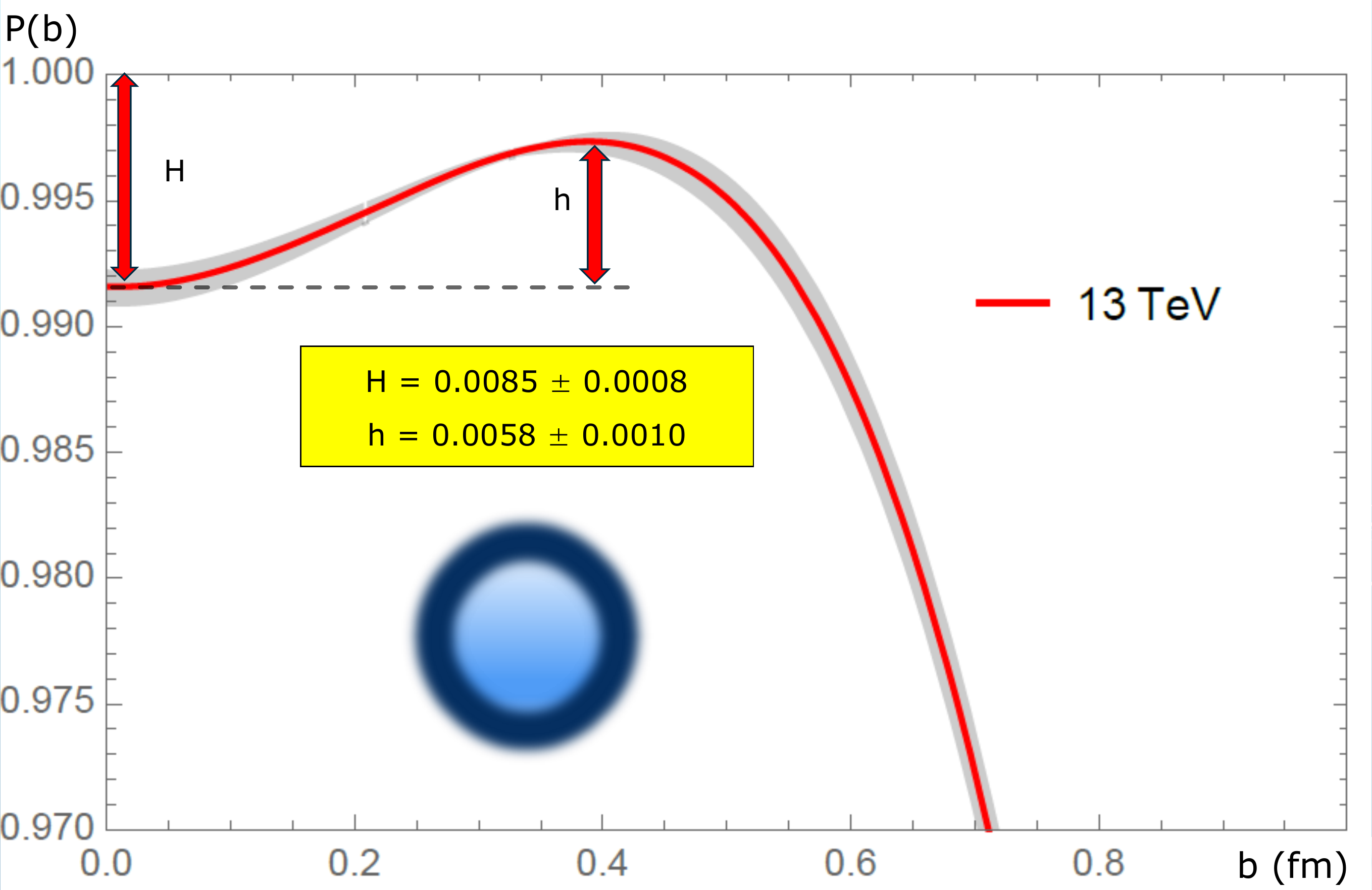}
        \caption{(Color online) The inelasticity (or shadow) profile function extracted from the TOTEM 13 TeV data 
        using the L\'evy expansion method, together with the associated error bar, zoomed in around the peak value. 
}
        \label{fig:hollowness}
\end{figure}
\begin{figure}[ht]
        \includegraphics[scale=0.68]{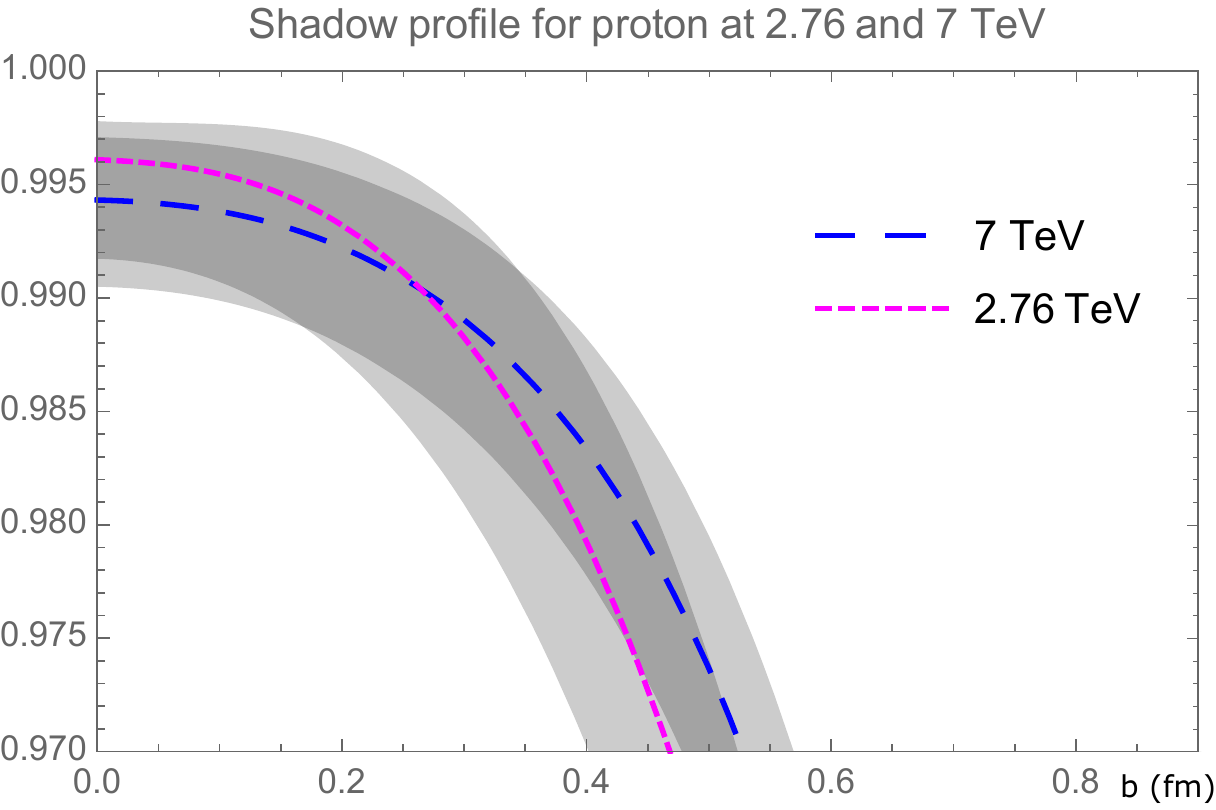}
        \caption{(Color online) The inelasticity (or shadow) profile function extracted from the TOTEM 7 TeV and 2.76 TeV data sets
        using the L\'evy expansion method, together with the associated error bar, zoomed in around the peak value. 
}
        \label{fig:hollowness-lower-E}
\end{figure}

At high enough energies, hollowness may thus become a generic property of the shadow profile functions, that characterizes the impact parameter distribution of inelastic scatterings. This effect contradicts the asymptotic black disc behaviour generally expected by the community~\cite{Block:2011vz} and thus it deals with a fundamental property of protons at asymptotically high energies.

We have carefully studied the small-$b$ region of profile functions for all existing data sets and found that such a hollow indeed appears in the proton, but only at currently highest LHC energy of 13 TeV. 
Fig.~\ref{fig:hollowness} zooms in the $P(b)$ inelastic scattering profile at 13 TeV in a vicinity of its peak value indicating the presence of such a hollow. This hollowness effect is a small but significant, much more than a 5$\sigma$ effect, characterized by a hollowness strength parameter of $h = 0.0058 \pm 0.0010$. The value of the depression at $b = 0$ is even more significant, it is characterized by $H = 1 - P(b= 0) = 0.0085 \pm 0.0008$ as illustrated on Fig.~\ref{fig:hollowness}. At 2.76 and 7 TeV, we do not find a significant hollowness effect, actually within large errors $h = P(b|{\rm max}) - P(b=0)$ is consistent with zero,
see Fig.~\ref{fig:hollowness-lower-E}. In a limited range of $\sqrt{s}$, the energy dependence of the elastic differential cross section satisfies certain scaling property recently found in Ref.~\cite{Csorgo:2019ewn}.
However, such a scaling is broken between 7 and 13 TeV i.e. at energy scales where the hollowness effect becomes significant. In order to make a reliable extrapolation of the hollowness towards the higher energies where the scaling property is violated, we need at least two points in this new scattering domain. For this purpose, a new TOTEM measurement at 14 TeV that is expected during Run 3 of the LHC will be highly relevant.

Our findings presented above are driven by the most recent and precise data from the LHC and by the power of the L\'evy imaging technique. These results may have a  profound impact on our theoretical understanding of the proton structure. Apparently, we found the first, statistically significant result, that suggests that at energies of $\sqrt{s} = 13$ TeV, the protons start to interact like  black rings rather than conventionally assumed black discs.

{\it Acknowledgments}
T. Cs. thanks to R.J. Glauber for inspiring discussions and for proposing a series expansion to describe elastic scattering. We acknowledge inspiring discussions 
with S. Giani, W. Guryn, G. Gustaf\-son, L. L\"onnblad, F. Nemes, K. \"Osterberg and M. \v{S}um\-bera and with members of the TOTEM Collaboration. 
R. P. is partially supported by Swedish Research Council grants No. 621-2013-4287 and 2016-05996, by ERC H2020 grant No 668679, as well as 
by the Ministry of Education, Youth and Sports of the Czech Republic project LT17018. T. Cs. and A. S. were partially 
supported by the NKFIH grants No. FK-123842, FK-123959 and K-133046, and by the  EFOP 3.6.1-16-2016-00001 grant (Hungary). 
Our collaboration was supported by THOR, the EU COST Action CA15213. 
 
\bibliographystyle{spphys}
\bibliography{biblio}

\end{document}